\documentstyle[pra,epsfig,aps]{revtex}
\input{psfig.sty}

\begin{document}
\pagestyle{plain}
\pagenumbering{arabic}
\title{Force and kinetic barriers to unzipping of the DNA double helix}

\author{Simona Cocco$^1$, 
R\'emi Monasson$^{2,3}$ and John F. Marko$^1$}
\address{$^1$ Department of Physics,
The University of Illinois at Chicago, \\ 845 West Taylor Street, Chicago
IL 60607-7059\\
$^{2}$ The James Franck Institute, The University of Chicago, \\
5640 South Ellis Avenue, Chicago IL 60637;\\
$^{3}$ CNRS-Laboratoire de Physique Th\'eorique de l'ENS, \\
24 rue Lhomond, 75231 Paris
cedex 05, France}
\date{\today}

\maketitle
\begin{abstract}
A theory of the unzipping of double-stranded (ds) DNA is presented,
and is compared to recent micromanipulation experiments. It is shown
that the interactions which stabilize the double helix and the elastic
rigidity of single strands (ss) simply determine the sequence dependent
 $\approx 12$ pN
force threshold for DNA strand separation. Using a semi-microscopic
model of the binding between nucleotide strands, we show that the
greater rigidity of the strands when formed into dsDNA, 
relative to that of isolated strands, gives rise to a potential
barrier to unzipping. The effects of this barrier are derived 
analytically.  The force to keep the
extremities of the molecule at a fixed distance, the kinetic rates
for strand unpairing at fixed applied force, and the rupture force as a
function of  loading rate are calculated.  
The dependence  of the kinetics and of the rupture force on molecule 
length is also analyzed.
\end{abstract}

%\vskip 10pt
 
{\bf Introduction:}
In cells, proteins apply forces to unzip and stretch DNA.  These
forces can be studied in single-molecule experiments
(Fig.~\ref{dessin})\cite{Ess,Leg,Lee,Rie,Stru,Str,Bon98}, and are of
biophysical as well as biological interest.  Our focus here is
primarily on unzipping experiments where forces are applied across the
double helix to adjacent 5' and 3' strands
(Fig.~\ref{dessin}A)\cite{Ess,Rie,Thom,Lub,Cac}.
In experiments, the control parameters may be the force $f$ itself,
the distance between the last base pairs $2r$, or the rate of 
force increase or `loading rate' (Fig.~\ref{dessin}B). 
We discuss the results expected in all these situations.
 
We first use a thermodynamical equilibrium approach to show that the
sequence-dependent force associated with unzipping of large DNAs,
$f_u \simeq 12 $ piconewton (pN),
can be simply deduced from the known free
energy of DNA denaturation and the elasticity of 
single-stranded DNA.  
The unzipping experiments of Essevaz-Roulet {\em et al}
(Fig.~\ref{dessin}A) \cite{Ess} and Rief {\em et
al} (Fig.~\ref{dessin}D) \cite{Rie} are accurately described at this
macroscopic level.

Other experimentally observable aspects of unzipping can only be
investigated using a more detailed description of base-pairing interactions.
We therefore present a semi-microscopic model which accounts for 
hydrogen bonds and stacking interactions \cite{Pro,Pey,Cul,Coc99,Coc00}.
We show that a free-energy potential barrier 
originates from the greater range of conformational fluctuation of
DNA strands when isolated, relative to when they are bound together
to form dsDNA (Fig.~\ref{potb175}).  Our model can be investigated in detail 
and allows precise calculation of the effects of this barrier
for the initiation of unzipping and the kinetics of strand dissociation.

We compute the force necessary to keep apart the two extremities of
the DNA molecules at some distance $2\,r$, as well
as the shape of the opening fork (Fig.~\ref{force}).
Due to the potential barrier, this
force is much larger at small $r$ (and can reach some hundreds of pN)
than the asymptotic value $f_u$ at large $r$.  Analysis of unzipping
in thermal equilibrium at the high level of precision possible in AFM
experiments would allow unambiguous verification of this predicted
force barrier.

The barrier makes strand dissociation an activated process with
dynamics that can be analyzed using nucleation theory\cite{Lan69}. 
Unzipping starts with a transition `bubble' a few ($\leq 4$) bases
long (Fig.~\ref{dessinbulle}).
We calculate the free energy of this bubble, and
determine how the dissociation rate depends on
applied force and molecule length  (Fig.~\ref{temps}). Results
are compared to the experiments of Bonnet {\em et al.} \cite{Bon98}
and of P\"orschke \cite{Por}.

Extending Evans' theory for the breaking of single bonds \cite{Eva97}
to the case of a one-dimensional polymer \cite{Seb00}, we then
calculate the most probable rupture force when the
DNA molecule is subjected to a force which increases at a
constant `loading rate' (Fig.~\ref{loading}). The dependence of the
rupture force upon loading rate and molecular length 
could be quantitatively tested by AFM unzipping experiments;
these results also shed light on AFM DNA-stretching experiments of
Struntz {\em et al.}\cite{Stru} and of Rief {\em et al.}\cite{Rie}.

{\bf Thermodynamic Description of Unzipping:}
An unstressed double helix is stabilized against spontaneously 
dissociating into its two strands by the interaction free energy per 
base pair, which from a thermodynamic perspective we may take 
to be some average amount $g_0$.  Although dependent on sequence,
we may consider $g_0 = -1.4 k_BT$, the value determined from single molecule
experiments on an AT rich sequence in $\lambda$ phage \cite{Str}, 
as a reference for the free energy difference between dsDNA and
separated ssDNAs.  Our emphasis is on an
understanding of the free-energy balance in unzipping rather
than to study inhomogenous sequence effects\cite{Thom,Lub}.

In the presence of applied torque
$\Gamma$ and unzipping force $f$ (Fig.~\ref{dessin}B)
the free energy difference per base pair
between unzipped and base-paired DNA strands is
\begin{equation} \label{deltag}
\Delta g (\Gamma,f) = g_{\rm ssDNA} - g_{\rm dsDNA} 
=  -g_0 + \theta_0 \Gamma + 2\, g_{\rm s}(f) 
\end{equation}
When $\Delta g < 0$, opening is thermodynamically favorable.
The last two terms represent mechanical work done per base pair
unzipping the double helix.  In order, they are the work 
done by the torque ($\theta_0 = 2\pi/10.5$ is simply the change in
strand winding angle during conversion of dsDNA to separated strands),
and the stretching free energy of the unzipped single strands.  

The function $g_{\rm s}(f)$ in (\ref{deltag}) 
is the ssDNA stretching free energy
per base at fixed force.  The leading factor of 2 simply reflects the
fact that two bases of ssDNA are created for each base-pair of dsDNA
which is unzipped.  The ssDNA elastic behavior is complicated by
nucleotide-interaction effects \cite{Maier}, but 
experimental force-extension curves for $\lambda$ phage ssDNA in 150 mM
Na$^+$ are well described by a freely-jointed chain-like (FJCL) elastic
response for forces $>1$ pN, with Kuhn length $d= 15 $\AA ~ \cite{bus,Rie}.
The corresponding free energy for forces up to $\approx 20$ pN is
\begin{equation} \label{forfjc}
g_{\rm s}^{FJCL} (f) = 
- k_B T \;\frac{l_{ss}}{d}\; \log\left[\sinh(d \, f / [k_B T])
\over  d \, f / [k_B T] \right ]
\end{equation}
where the contour length per base pair is 
$l_{ss}=27$ $\mu$m$/48.5$ kb $\simeq 5.6$ \AA.

At zero applied force, $g_{\rm s}(0)=0$, thus ssDNA is stable when
$\Gamma < \Gamma_u = g_0/\theta_0 = -2.4 k_B T$, in good agreement
with an experimental estimate of the twisting torque needed to
denature an AT rich sequence in a $\lambda$ phage DNA \cite{Str}
(the sign indicates a left-handed dsDNA-unwinding torque).  In this
case, the work done by the torque during opening is simply $-g_0 = 1.4
k_B T$ per base pair.

In the opposite case where torque is zero ({\em i.e.} for dsDNA with no
constraint of its strand linking number), the critical unzipping force
(at which $\Delta g(f)=0$)
is $f_u ^{FJCL} = 11$~pN. These results are in good agreement with the
mid-range of unzipping forces encountered with experiments on
inhomogeneous-sequence DNAs by Essevaz-Roulet {\em et al.}
(Fig.~\ref{dessin}A: 12 pN threshold to start, then 10 to 15 pN during
unzipping $\lambda$-DNA\cite{Ess}). 
Data of Rief {\em et al.} 
(Fig.~\ref{dessin}D; $20 \pm 3$ pN for poly(dG-dC), $9 \pm 3$ pN for
poly(dA-dT)\cite{Rie}) gives via (\ref{deltag}) denaturation free energies of
1.1 and 3.5 $k_B T$ per AT and GC base pair respectively, in
good agreement with thermodynamical data\cite{Breslauer}.
Finally, the projected length of one ssDNA nucleotide along the
force direction at the unzipping transition is given as
$d_u^{FJCL}\simeq 4$\AA.
Bockelman {\it et al.} used a similar theory
to analyze unzipping force dependence on sequence\cite{Ess}.

The inset of Fig.~\ref{force} shows the curve in the torque-force plane on
which $\Delta g = 0$, which is the `phase boundary' separating dsDNA
and unzipped ssDNAs. This boundary is predicted to have the shape $f_u
\propto (\Gamma - \Gamma_u )^{1/2}$ for small $f_u$.  

For forces up to 15 pN, (\ref{forfjc}) is approximated to
$0.15 k_B T$ accuracy by the simple quadratic form
%$g ^{G}_{\rm s} (f) = - f^2/C$
\begin{equation} \label{forgau}
g ^{G}_{\rm s} (f) = - \frac {f^2}{C} \qquad .
\end{equation}
where the ssDNA effective elastic constant is $C=0.12 \, k_B T/$\AA $^2$.
Using this form allows analytical solution for the unzipping force,
$f_u ^{G}=( C \, |g_0|/2)^{1/2} =$ 12 pN \cite{Lub,Cac,Seb00};
at this force the projection of ssDNA monomer length 
along the force direction is $d_u^{G} = (2|g_0|/C)^{1/2} \simeq 5$\AA.
This quadratic approximation is quantitatively nearly equivalent to
the nonlinear model (\ref{forfjc}) at forces up to $\approx 15$ pN
(e.g. note the accord between the torque-force `phase boundaries' 
in Fig. 3, inset);
this will be key to the continuum theory below.

{\bf Semi-Microscopic Model of Strand Binding:}
Features at the nucleotide scale relevant to the onset of unzipping are
ignored in models like (1).  We therefore move to a model which uses
the distances $2\,r(n)$ between corresponding $n$th base pairs of 
the two strands, as degrees of freedom.  The energy of the DNA strands is:
\begin{equation} \label{ham}
H = \int_0^N dn \left\{ {1 \over 2}  m \big( r(n) \big) \left( dr \over dn 
\right)^2 + U\big( r(n) \big) \right\}
\end{equation}
This model is similar to models previously used to describe
thermal denaturation\cite{Pey,Cul}.  The precise form of (\ref{ham})
follows from our previously developed model for denaturation by 
torque \cite{Coc99,Coc00} by integration over angular degrees of
freedom, followed by continuum limit for the base index $n$.

The first term in (\ref{ham}) describes interactions between neighboring 
bases along each strand, and must depend strongly on the 
inter-base half-distance $r(n)$, since conformational fluctuations of
the strands are highly quenched inside the double helix, relative to those
along ssDNA.  The strand rigidity is
$m(r) = E\, e^{-b(r-R_0)} + C$, 
where $R_0 = 10$ \AA ~ is the double
helix radius, and where $1/b=0.6$ \AA ~ is the separation at which 
the strand rigidity changes from its double-helix value $E+C$, to the much
smaller ssDNA value $C$.  We use the value $C = 0.12$ $k_B T/$\AA$^2$
from the previous section.
The rigidity of the strands inside the double helix has been 
determined from Raman measurements of internal vibrations of
dsDNA, to be $E = 58 k_B T/$\AA$^2$\cite{Coc00,Ura81}. 

The second term in (\ref{ham})  is a  potential, acting between
the two strands, made up of the
hydrogen-bonding energy between corresponding bases,
plus a torque energy:
$U(r) = U_H (r)  - \Gamma R_1/r$
where $R_1 = 6$ \AA\cite{Coc99}.
We use the Morse potential form\cite{Mor} for the 
hydrogen bonding interaction\cite{Pro,Pey}:
$U_H(r) = D [ (e^{-a(r-R)} - 1)^2 - 1]$, with $D=5.84 k_B T$ and
$a = 6.3$ \AA$^{-1}$ (Fig.~\ref{potb175} inset).

In thermal equilibrium, strand unpairing is described by the 
partition function
$Z = \int {\cal D}r(n) ~ e^{-H/k_B T}$. 
$Z$ can be computed by use of a continuum transfer
matrix technique along the $n$ coordinate, leading to a Schr\"odinger-like
equation:
\begin{equation} \label{sc}
\left[ -{(k_B T)^2 \over 2\, m(r)} \, \frac{\partial ^2}{\partial r ^2} 
+ V(r) \right] \psi(r) = g\, \psi(r) \qquad .
\end{equation}
The free-energy potential $V(r)=U(r) +  (k_B T/2)\ln[m(r)/m(\infty)]$
includes an entropic contribution due to the
decrease in the rigidity $m(r)$ with strand unbinding.
This entropic potential (Fig.~\ref{potb175}) arises 
when going from the path integral
to the Schr\"odinger equation \cite{deg} with non constant mass\cite{Cul},
and here it generates a large force barrier to the 
initiation of unzipping, which strongly affects the kinetics 
of strand separation.

The lowest eigenvalue $g_0$ is the equilibrium free energy, and the
corresponding $\psi_0(r)$ describes thermal fluctuations of distance
between the two strands.  For $g\geq 0$ the eigenvalue
spectrum becomes continuous, corresponding to the appearance of
completely separated ssDNAs at $g=0$. When DNA is the thermodynamical
favorable state $g_0 < 0$ represents the free energy per base pair of
a long dsDNA, relative to separated ssDNAs.  The fluctuations of $r$
are confined to the Morse well, so $g_0$ is well approximated if we
take $V(r) = V(R_0)+ U_H(r)-U_H(R_0)$ and $m(r)=m(R_0)$ for which 
(\ref{sc}) is exactly soluble\cite{Mor}.  Inside the well, 
$\psi _0(r)$ is the Morse ground state \cite{Mor}; outside the well it
can be computed using the WKB approximation\cite{Boh}.
Using the parameters listed above, and at zero torque, the lowest
eigenvalue is $g_0 = -1.4 k_B T$, in accord with the corresponding
number assumed in the previous section.  Application of an unwinding
torque gradually increases $g_0$ until at $\Gamma_u = -2.4 k_B T$ it
becomes zero and the DNA unwinds, exactly as occurs in the
thermodynamic model of the previous section (Fig. 3, inset).

{\bf Force required to hold ssDNA ends at a given distance}:
This situation can be analyzed simply
in terms of $\psi_0(r)$, without further computation.  This is because,
for a semi-infinite dsDNA, the function $\psi_0(r)$ corresponding to
$g_0$ is the probability distribution for the two ssDNA ends to fluctuate
a distance $2r$ apart.  Therefore the free energy associated with a 
fluctuation which separates the two ssDNA ends by a distance $2r$ (or
equivalently the total work done separating the two ends to a distance 
$2r$) is
\begin{equation}
W(r) = -k_B T\log\psi_0(r)
\end{equation}
up to an additive constant which is unimportant for our analysis.
%In a rough approximation consisting in
Note that, neglecting surface interactions,
 this is  half the free energy associated 
to separate a distance $2r$ in the middle of a long dsDNA,
since an interior `bubble' is made of two `forks'.

The unzipping force that must be supplied to hold the two ssDNA ends a
distance $2r$ (Fig.~\ref{dessin}B)
apart is thus just the derivative of (6), $f(r) =
dW(r)/d(2r)$.  This is in Fig.~\ref{force}, which displays a large
force barrier of $\approx 270$ pN as the strands are forced apart.
The barrier peak occurs for a half-separation $r-R_0 = 0.5$ \AA, and
then decays to the long-molecule unzipping force 
$f_u \simeq 12$ pN by $r-R_0 \approx 4$ \AA.  
At large distances, the work done per base pair by unzipping is
$2 f_u d_u \simeq 3 k_B T$,  twice as much as the denaturation free
energy $-g_0$; this is because the force 
must unzip the DNA, and extend the highly flexible ssDNAs.
The peak force is large compared to the fluctuations in force 
associated with sequence\cite{Ess}.
The force barrier will not be observable in large-scale unzipping
experiments\cite{Ess}, but should be observable in AFM studies.  A
stiff cantilever with roughly 0.1 \AA ~ thermal noise should be used
to measure the force barrier as a function of essentially fixed
opening distance.

In this fixed-distance experiment, one might
also measure the shape of the opening `fork', by determining
the relation between opening distance and base position,
$n(r)$ (Fig.~\ref{dessin}B). 
% There is of course a probability distribution
%for $n(r)$, but because the forces needed to unzip dsDNA are
%sufficient to appreciably stretch out ssDNA (see previous section),
%this distribution is well described by its peak.  
The most probable configuration $n(r)$
satisfies the equation of motion associated with (\ref{sc}),
which expresses force balance along the chain,
% including the
%effect of harmonic fluctuations (i.e. including the entropic potential
%of (\ref{sc})
\begin{equation} \label{classic}
m(r) {d^2 r \over dn^2} + \frac 12 {m'(r)} 
\left( dr \over dn \right)^2 = V'(r)
\end{equation}
We integrate (\ref{classic}) to obtain $r(n)$,
the shape of the opening `fork'. 
Starting from the opening point where
$r=R_0$ and $dr/dn=0$, $n(r) = 2 [E\,(r-R_0)/V'(R_0)]^{1/2}$.  Far from
the opening point $n(r) = r/d_u^{G}\simeq r/(5$\AA )\cite{Cac}.

{\bf Unzipping kinetics at fixed force:}
Many experiments on short (10-100 bp) dsDNAs 
(\cite{Lee,Stru,Bon98,Por} and caption of Fig.~\ref{dessin})
probe the kinetics of strand separation.
The equilibrium results discussed above are a starting point for
a kinetic theory of unzipping based on nucleation theory\cite{Lan69}.
The Schr\"odinger equation (\ref{sc}) with the
fluctuation-corrected potential $V(r)$ describes dsDNA and ssDNA as 
locally stable molecular states. 
The general problem faced in unzipping kinetics is the transition
from an initially metastable state 
(dsDNA or ssDNA, depending on the force, see below)
to a final, stable (lower-free-energy) state.
Strand dissociation requires the whole polymer chain
to cross the free energy potential barrier of $V(r)$ (Fig. 2), which
makes the transition rate strongly length and force dependent.
Using the effective potential $V(r)$ (including the entropic
barrier) corresponds to averaging over microscopic fluctuations 
of individual bases, restricting us to consider unzipping rates slow 
compared to those of these microscopic fluctuations (the
experimentally relevant regime).

The transition rate (equivalently the inverse lifetime) has the
form familiar from transition-state theory \cite{Lan69}:
\begin{equation} \label{rate}
\nu = \nu_0 \; e^{-G^*/k_B T}
\end{equation}
It requires the activation free energy $G^*$ of a transition state, 
relative to the initial metastable state.
The transition state is the saddle-point configuration of the
free energy, with one unstable direction leading monotonically 
down to the initial and final states, and is the dominant
transition pathway\cite{Lan69}.

Our transition states are just the partially unzipped 
configurations $r^*(n)$ determined from (\ref{classic}), for force-dependent
boundary conditions consistent with the initial
($r_i$) and final ($r_f$) states.
The activation free energy is 
\begin{eqnarray}
\label{activation}
G^* &=& H[r^*(n)] + (k_B T/2)\int dn \ln[m[r^*(n)]/m(\infty)] 
- 2 f r^*(0) - G_{m}  \nonumber \\
 &=&\int _{r_{i}} ^{r_{f}} dr \, \sqrt{ 2 \, m(r) \, (V(r)-g_{m})}
\ - \ 2\, f \,(r_{f}-r_{i}) \qquad ,
\end{eqnarray}
i.e. the free energy associated with (\ref{classic}) minus the
free energy of the metastable state $(G_{m}= N g_{m})$ 
from which the transitions occur.
$G^*$ is in practice the free energy of the
few-base-pair `bubble' portion of the transition state which
separates the unzipped and double-stranded regions.

Finally, the rate prefactor $\nu_0$ is the linear growth rate of
unstable perturbations around the saddle-point configuration\cite{Lan69}.
We assume viscous dynamics with a friction coefficient per base
$\zeta = 6\pi\eta R_0$
(water viscosity $\eta = 1\times 10^{-3} ~ {\rm kg}/({\rm m}\cdot 
{\rm sec})$).  From a detailed calculation we find 
$\nu_0 \approx Da^2/(4\zeta) = 1\times 10^{12}$ sec$^{-1}$,
essentially the ratio of the negative curvature of the
Morse potential near the top of the well, to the friction coefficient.
We now describe how to compute the dissociation rate 
and the nucleation bubble shape (i.e. boundary condition $r_i, r_f$)
depending on the unzipping force.
  
{\bf Kinetics of unzipping where ssDNA is stable ($f > f_u$):}
If a steady force is applied which is slightly bigger than the equilibrium
unzipping threshold $f_u$, then the initial dsDNA is metastable
relative to separated strands: $\Delta g(f)=-g_0-2 f^2/C <0$  (\ref{deltag}),
 and formula (\ref{rate}) 
is directly applicable (Fig.~\ref{dessinbulle}A) to calculation of the 
dissociation rate $\nu_{-}$. 
The free energy of the initial metastable state is $g_m=g_0$.
The transition state in this case is a short
ssDNA `nucleus' of $n^*$ bases at the open end ($n=0$) of the dsDNA,
with boundary conditions at the ends 
of fixed force $f$ ($dr/dn = 2 f/m[r(n)]$),
and free energy equal to the metastable dsDNA value 
($V(r(n)) - 2 f^2/m[r(n)]= g _0$).
This determines $r_{i}=r (n^*)$ and $r_{f}=r(0)$.

The size of the nucleation `bubble' depends weakly
 on the force and is $n^*(f) \simeq 4$ bases, independent
of the overall DNA molecule length.
Therefore, the dissociation
time $t_- = 1/\nu_-$ (Fig.~\ref{temps}) is length-independent
for forces above $f_u = 12$ pN.  
The decrease in dissociation time with forces $> f_u$ is
due to reduction of $G^*$ by the applied force.
Beyond $f_b\simeq 230$~pN, the barrier in $V(r)$ is completely overcome,
and unzipping is immediate ($t_{-}=1/\nu_0$). 
Our computation addresses only the initial unzipping
barrier-crossing event and does not include the time necessary to push
the fork down the dsDNA, which would introduce a weak molecular-length
dependence for forces $> f_u$ in Fig.~\ref{temps}.
Fork motion near $f_u$ may also be retarded by additional barriers
introduced by inhomogeneous sequence.
{\bf Kinetics of unzipping where ssDNA is metastable ($f < f_u$):}
Below the equilibrium unzipping force threshold, an infinitely long
molecule is stable as a dsDNA.  However, relatively short 
($<20$ bp) oligomers have a finite
strand-dissociation time which can be accelerated by applied force.
Here the final ssDNA state is metastable (although with a possibly
long lifetime) relative to the initial dsDNA state.   
Therefore our calculation scheme applies not to 
unzipping (off-rate $\nu_-$), but instead to {\it annealing} of ssDNAs 
(on-rate $\nu_+$, Fig.~\ref{dessinbulle}B).
We compute the off-rate from the on-rate, using the
equilibrium condition $\nu_-/\nu_+ = e^{-N \Delta g(f)/k_BT}$,
where $N$ is the number of base-pairs in the molecule, and where
$\Delta g(f)$ is the force-dependent free energy difference per base pair
between the paired and unpaired states .

The calculation of the on-rate requires computation of a saddle point
configuration with a nucleation bubble of dsDNA at the end of ssDNAs
to which the force is applied (Fig.~\ref{dessinbulle}B).   
The relevant boundary 
conditions are analogous to those above, apart from the requirement
that the ssDNAs be in close proximity but not base-paired (i.e. slightly
outside the potential barrier of Fig.~\ref{potb175}) and the
free energy of the metastable initial state equals $ g_m=2\,g_s(f)= - 2 f^2/C$.
The dsDNA nucleation `bubble' remains near 4 bp
as the force $f$ decreases from $f_u$ to zero, leading to an
essentially $N$-independent on-rate $\nu_+$ (again the time necessary
for the actual `zipping up' is not included).   
Then, the $N$-dependence from the energy difference of paired and unpaired
DNA results in a strong molecular-length dependence of $\nu_-$
as shown in Fig.~\ref{temps} for forces less than $f_u$.

At zero force, this becomes a calculation of
dissociation time for free dsDNAs in solution with result
$t_- = 10^{0.6 N - 6.3}$~sec.  The inset of Fig.~\ref{temps} shows
this as a function of dsDNA length; a 10 bp dsDNA has a lifetime of
roughly 1 sec; the exponential length-dependence results in a 30 bp
DNA being stable for $\approx 10^{12}$ sec$\approx 30000$ yr.
Our estimate of $t_-$ for $N=5$, $t_-\approx 0.3$ msec, is in agreement 
with the results of Bonnet {\em et al.} \cite{Bon98} (Fig.~\ref{dessin}E).
The prediction for the nucleation bubble size, $n^*=4$, is close to
the value $n^*=3$ measured by P\"orschke for a poly(A)poly(U) acid\cite{Por}.

{\bf Rupture of dsDNA during gradual loading:}
The previous two sections discuss fixed-force experiments; an alternate
experiment is to steadily increase force ($f = \lambda t$ where 
$\lambda$ is the `loading rate' in pN/sec) 
and then to measure the force at which rupture occurs.
Generally, rapid loading rates result in rupture at 
large forces.  We are able to predict the most probable rupture force 
versus loading rate and molecular length, and our results display a
rich range of possibilities (Fig.~\ref{loading}).

Using the calculations of off-rates presented above, the probability
distribution for rupture with force is\cite{Eva97}
\begin{equation}
P(f) = \nu_-(N,f) \; \exp \left( 
- {1 \over \lambda} \int_0^{f} df' \nu_-(N,f') \right)
\end{equation}
For a number of molecular lengths,
Fig.~\ref{loading} shows the location of the peak of this distribution,
the most probable rupture force that would be measured experimentally.
For sufficiently slow loading rate, rupture occurs at
zero force simply by thermal dissociation; for molecules $> 20$ bp
thermal dissociation is practically unobservable.

At some loading rate $\lambda_1$, the peak in $P(f)$ jumps to
finite force and then increases, with each length of molecule
following a different curve.  In this regime, the forces at
rupture are typically below the equilibrium unzipping 
threshold $f_u$, since there is time for many thermal
attempts at barrier-crossing to the metastable ssDNA state during
loading.  The length-dependence follows from the calculation of formation
of metastable ssDNA discussed above, and has been qualitatively
observed for 5'-5' pulling experiments (Fig.~\ref{dessin}C)\cite{Stru}. 

At a higher loading rate $\lambda_2$ the peak of the rupture force
distribution hits $f_u$ which remains the most likely
rupture force up to a loading rate $\lambda_3$.
For $\lambda > \lambda_2$, dissociation is  occuring at forces large enough 
that the ssDNA final state is stable, resulting in no $N$-dependence. 
Finally, beyond the very large loading rate 
$\lambda_3 \approx 10^{5.5}$ pN/sec, the
rupture force gradually increases simply because the molecule is
unable to respond to the force before it becomes very large.
$\lambda_3$ separates equilibrium and nonequilibrium time scales for very 
long sequences ($N\to \infty$): the rupture force is independent on the 
loading rate 
and equal to $f_u$ when $\lambda < \lambda _3$, and increases above.

Our framework treats nonequilibrium rupture of a one 
dimensional object, a development of previous theory\cite{Eva97}
necessary for interpretation of unzipping experiments.
In light of our detailed predictions, 3'-5' AFM unzipping experiments
(Fig.~\ref{dessin}A) should yield interesting results, and would be
of help for an accurate determination of the free-energy potential $V(r)$.

We thank C. Bouchiat, G. Bonnet, V. Croquette, F. Pincet for useful 
discussions. Work at UIC was supported by NSF Grant DMR-9734178, 
by the Petroleum Research Foundation of the American Chemical Society 
and by the Research Corporation.
S. Cocco is partly funded by A. della Riccia grant. R. Monasson is
supported in part by the MRSEC Program of the NSF under Award number
DMR-9808595.

\begin{center}
$^{\, }$ \vskip 2cm
\begin{figure}
{\bf A}
\includegraphics[height=140pt,angle=0] {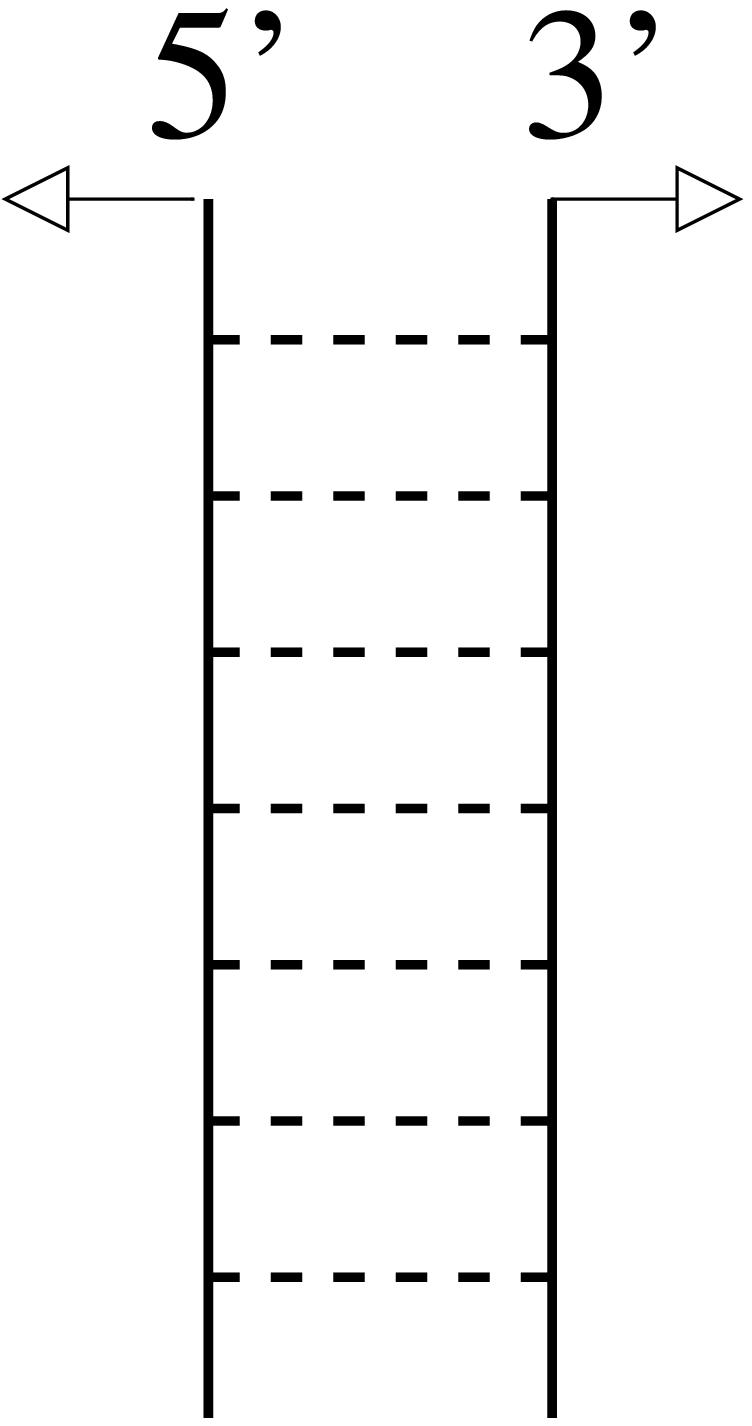}
\hskip 4cm
{\bf B}
\includegraphics[height=160pt,angle=0] {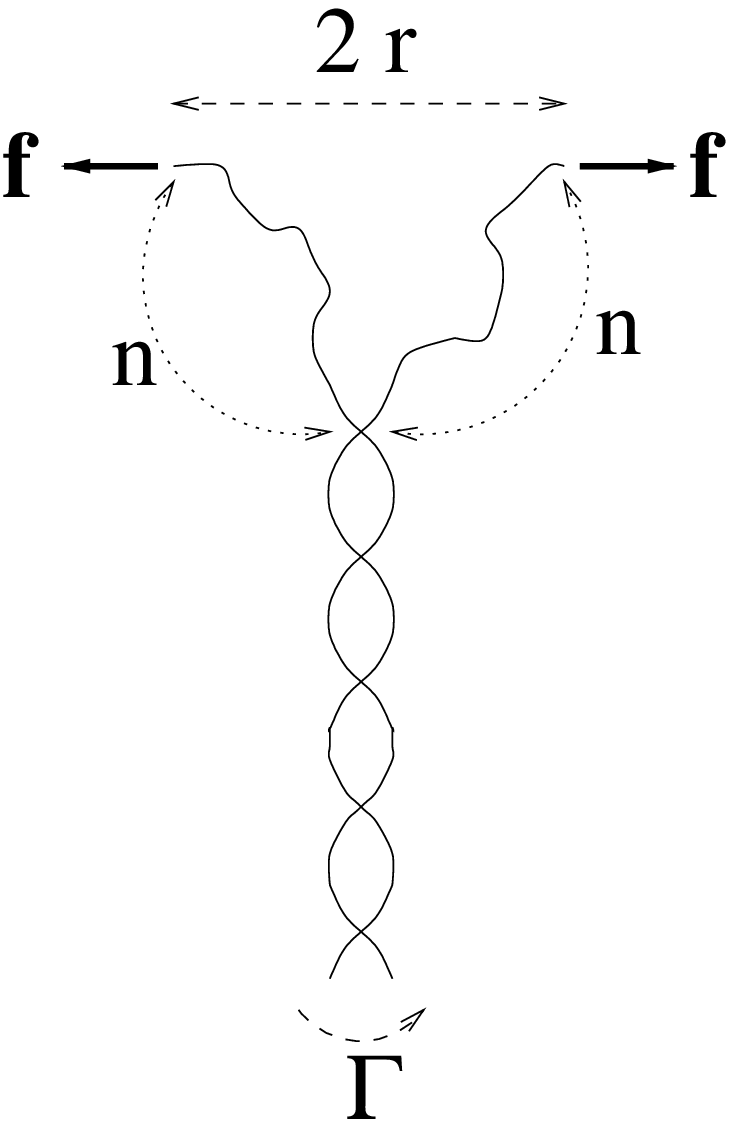}
\hskip 8cm \vskip 2cm
{\bf C}
\includegraphics[height=140pt,angle=0] {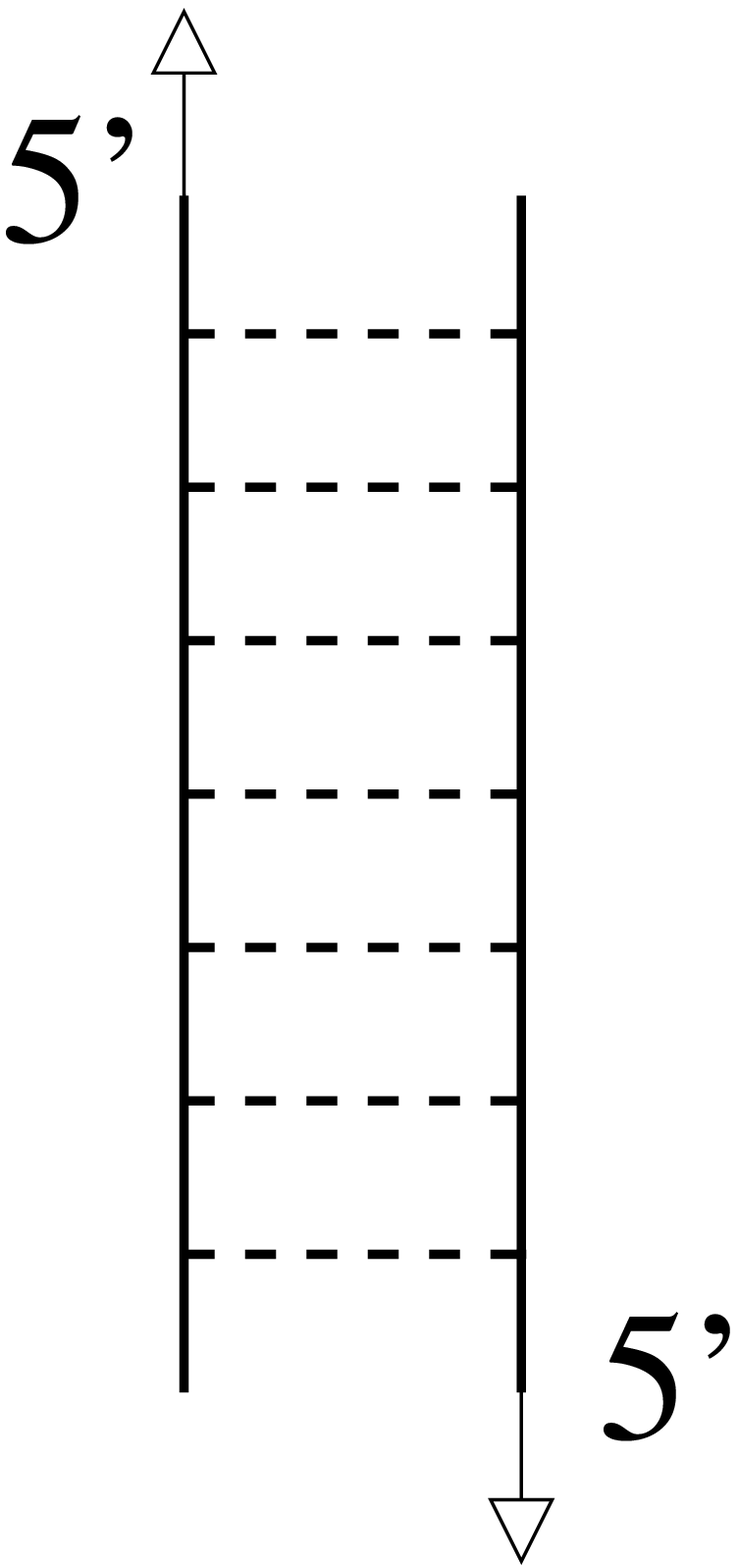}
\hskip 4cm
{\bf D}
\includegraphics[height=140pt,angle=0] {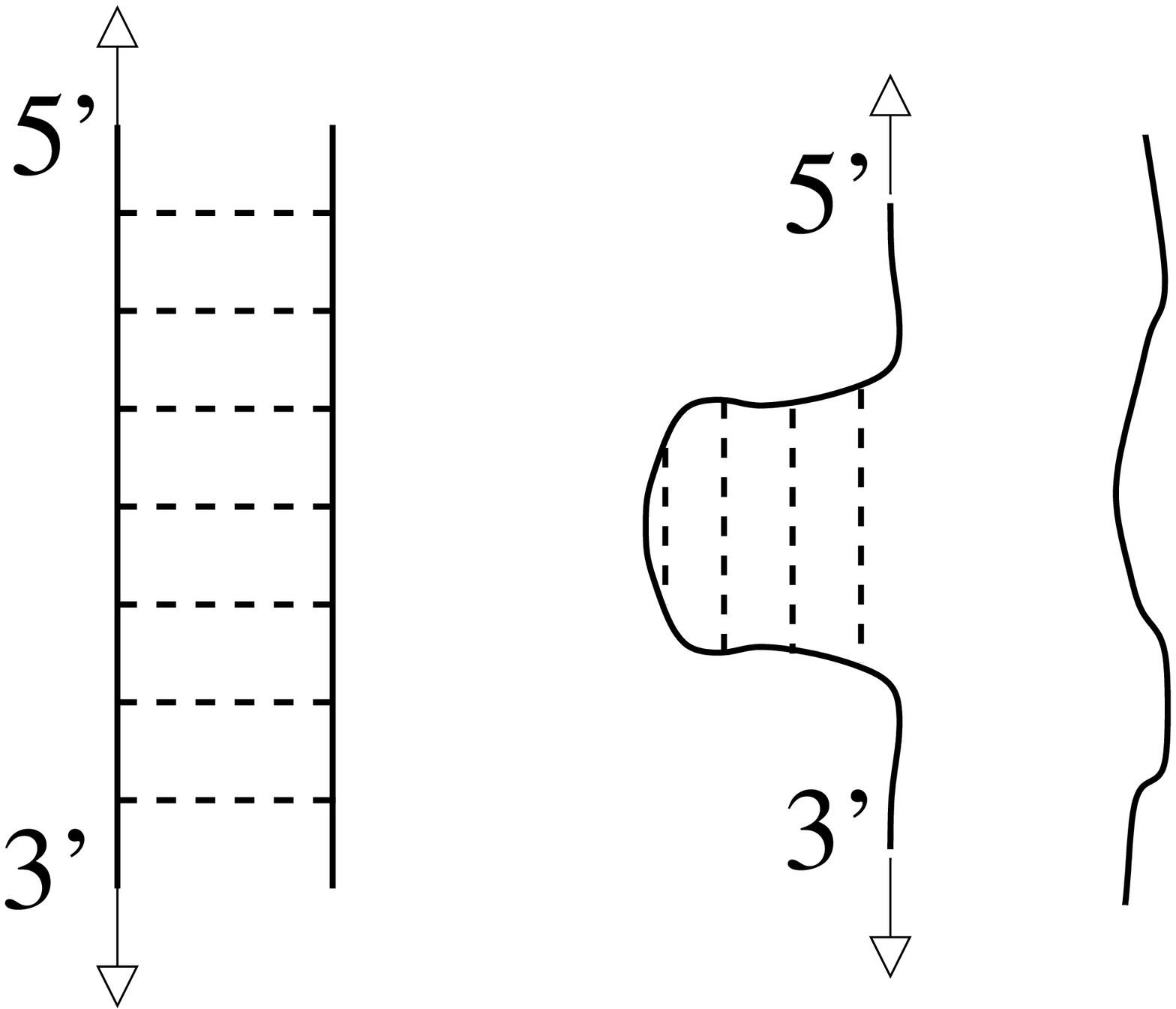}
\hskip 8cm \vskip 2cm
{\bf E}
\includegraphics[height=140pt,angle=0] {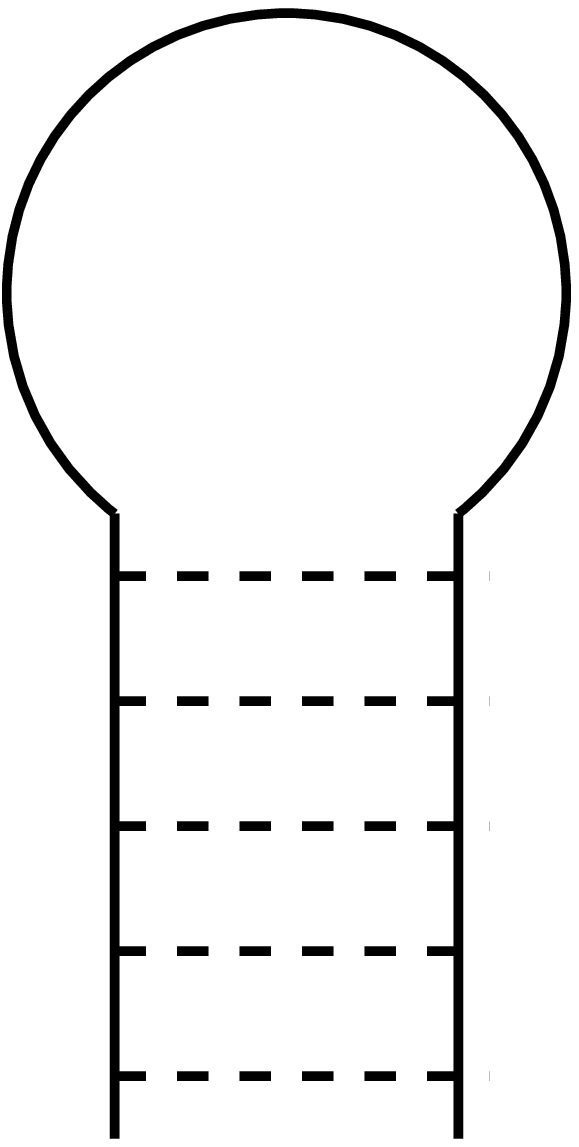}
\vskip 0.5 cm
Cocco. Fig~1. Desired size: 2 columns, heigth 6.5 cm. 
\vskip 1cm
\caption{Sketches of some experiments referred to in the text. 
All experiments are at room temperature and in physiological liquid
buffers (PBS or Tris). Arrows symbolize the applied forces.  {\bf A}:
Unzipping experiment of Essevaz-Roulet {\em et al.}  [1]: the
3'-5' extremities of a $\lambda$-phage DNA (49 kbp) are
attached to a glass microscope slide (with translational velovity $v
=40$ nm/sec) and a polystyrene bead connected to a glass microneedle
(with stiffness $k=1.7 $~pN~/$\mu$m).  The loading rate equals $
\lambda = k v= 0.06$ pN/sec.  When the force approaches 12 pN, the DNA
starts to open. As unzipping proceeds, the distance between the two
 single strands extremities is controlled and 
the  force varies between 10 and
15 pN depending on the sequence.  
{\bf B}: parameters used in the theoretical description: force $f$, 
torque $\Gamma$ and
distance $2\, r$ between the two single strands extremities.
{\bf C}: Stretching experiment of
Strunz {\em et al.}[5]: a short ssDNA (10, 20 or 30 bp with
about 60\% GC content) is attached by one 5'-end to a surface, the
complementary ssDNA is attached by the other 5'-end to an AFM tip.  On
approaching of the surface to the tip, a duplex may form that is
loaded on retract until unbinding occurs.  The distribution of the
rupture forces is obtained for loading rates ranging from 16 to
4000~pN/sec.  {\bf D}: Stretching and unzipping experiment of Rief
{\em et al.}[4]: DNA of poly(dA-dT) (5100 bp) or poly (dG-dC)
(1260 bp) are attached between a gold surface and an AFM tip, and
stretched.
Through a melting transition, single DNA strands are prepared; these
strands upon relaxation reanneal into hairpins as a result of their
self-complementary sequences. The forces of unzipping of these
hairpins are $20 \pm 3$~pN for poly (dG-dC) and $9 \pm 3$~pN for
poly(dA-dT).  {\bf E}: Dissociation experiment of Bonnet {\em et
al.}  [7]: The rate of unzipping, $\nu _-$, and closing,
$\nu_+$, of a 5 bp DNA hairpin (CCCAA-TTGGG) is investigated by
fluorescence energy transfer and correlation spectroscopy
techniques. The hairpin is closed by a loop of 12 to 21 Thymine (T) or
Adenine (A).  The characteristic time of opening $t_-=1/\nu_-$ is
found to be largely independent of the loop length, and equal to
$t_-\simeq 0.5$~msec.}
\label{dessin}
\end{figure}
\end{center}

\begin{center}
\begin{figure}
\includegraphics[height=350pt,angle=-90]{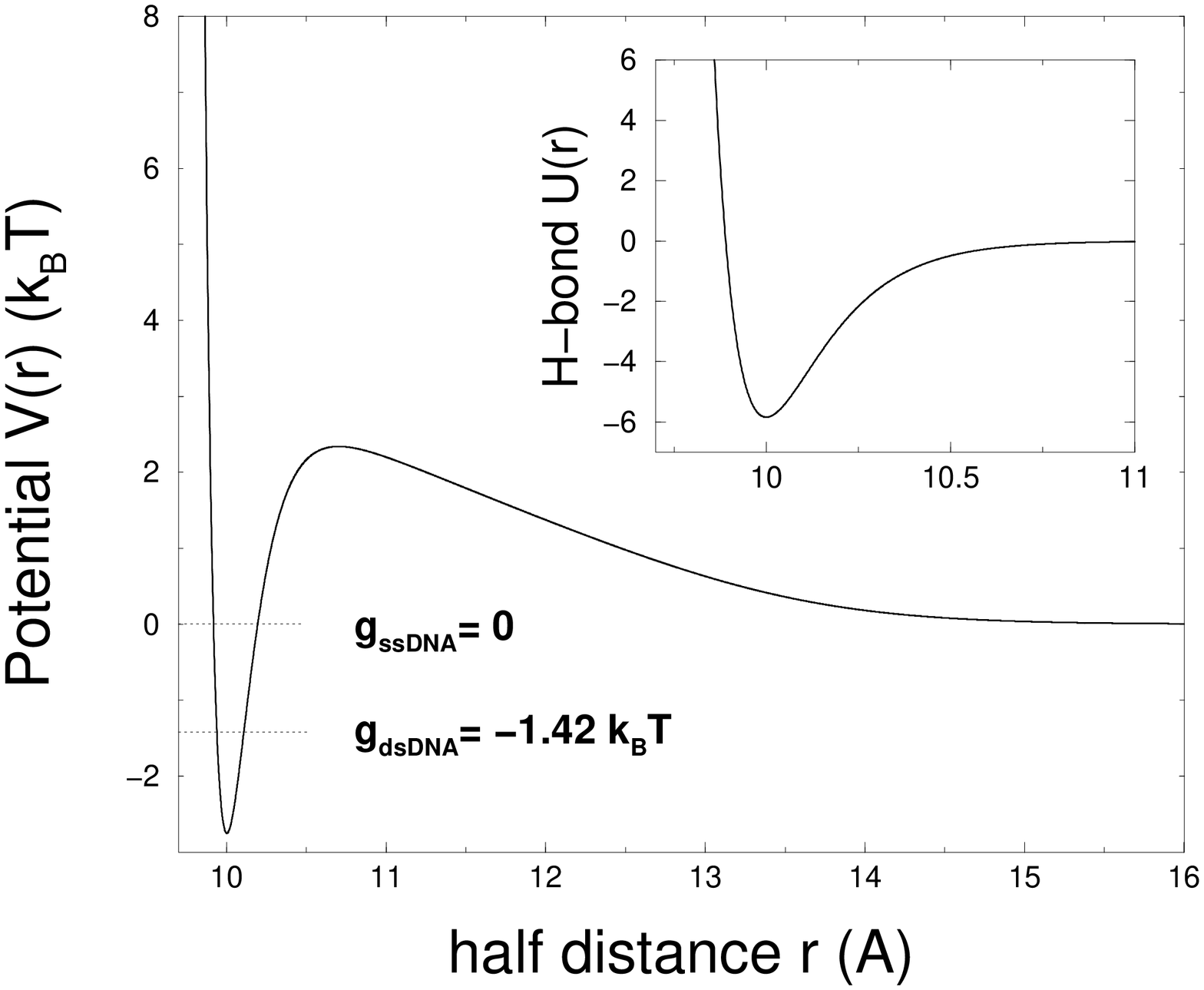}
\vskip 0.5 cm
Cocco. Fig.~2. Desired sized: 1 column, height 8.5 cm.
\vskip 1.5cm
\caption{Base pair potentials in unit of $k_BT$, as a function of the
base radius $r$ (in \AA), without (inset) and with (main picture)
entropic contributions. Inset: Morse potential $U(r)$ accounting for
the hydrogen bond interaction. Main picture: total potential $V(r)$
for zero torque. Once entropic contributions are considered, small $r$
values are less favorable and a barrier appears. The free-energy
$g_{dsDNA}=g_0$ of the dsDNA is lower than the single strand free-energy
$g_{ssDNA}=0$.  Note the difference of scales on the horizontal axis
between the two figures.}
\label{potb175}
\end{figure}
\end{center}

\begin{center}
\begin{figure}
\includegraphics[height=350pt,angle=-90] {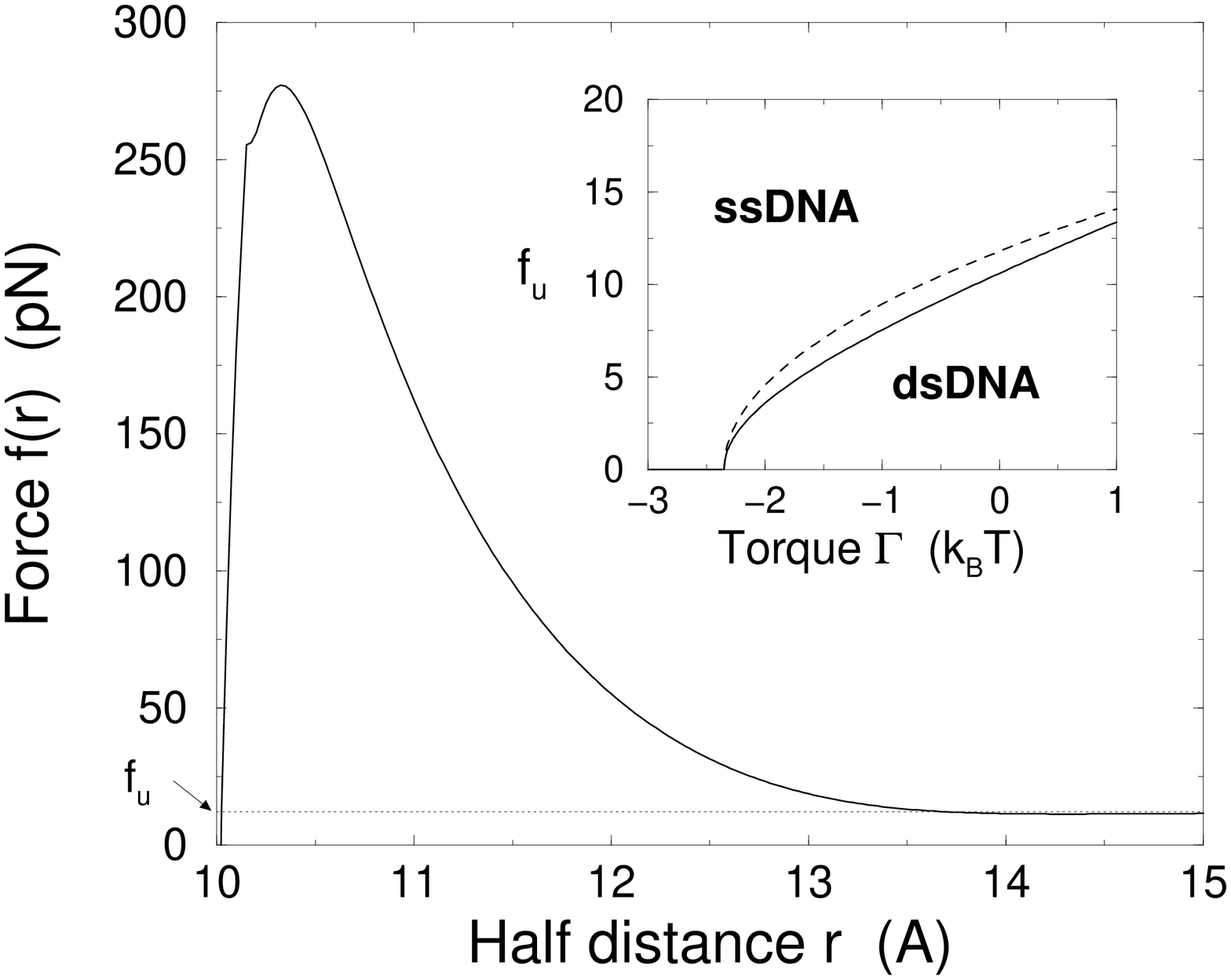}
\vskip 0.5 cm
Cocco. Fig~3. Desired size: 1 column, height 8.5 cm.
\vskip 1.5cm
\caption{Force $f(r)$ (in pN) to be exerted on the DNA to keep extremities at
a distance $2\, r$ apart (in \AA). The peak force $f \simeq 270$~pN, reached at
$r\simeq 10.5$~\AA, is much larger than the asymptotic
value $\simeq 12$~pN, equal to the equilibrium force $f_u$ (at zero torque,
and in the Gaussian approximation)
for unzipping a large portion of the molecule.
Inset: phase diagram, in the plane of torque $\Gamma$ ($k_B T$)
and of force $f$ (pN). The lines shows
the critical unzipping force $f_u$ as a function of $\Gamma$
 with formula (2) (full line) and formula (3) (dashed line) for
the stretching free energy of the single strand. Below
the line, dsDNA is the stable thermodynamical configuration while for forces
larger than $f_u (\Gamma)$, denaturation takes place. $f_u$ vanishes at the
critical torque $\Gamma _u \simeq -2.4 \; k_BT$, and is equal to
11 pN (full line) and 12 pN (dashed line) at zero torque.}
\label{force}
\label{diag}
\end{figure}
\end{center}

\newpage 
\begin{center}
\begin{figure}
{\bf A}
\includegraphics[height=140pt,angle=0] {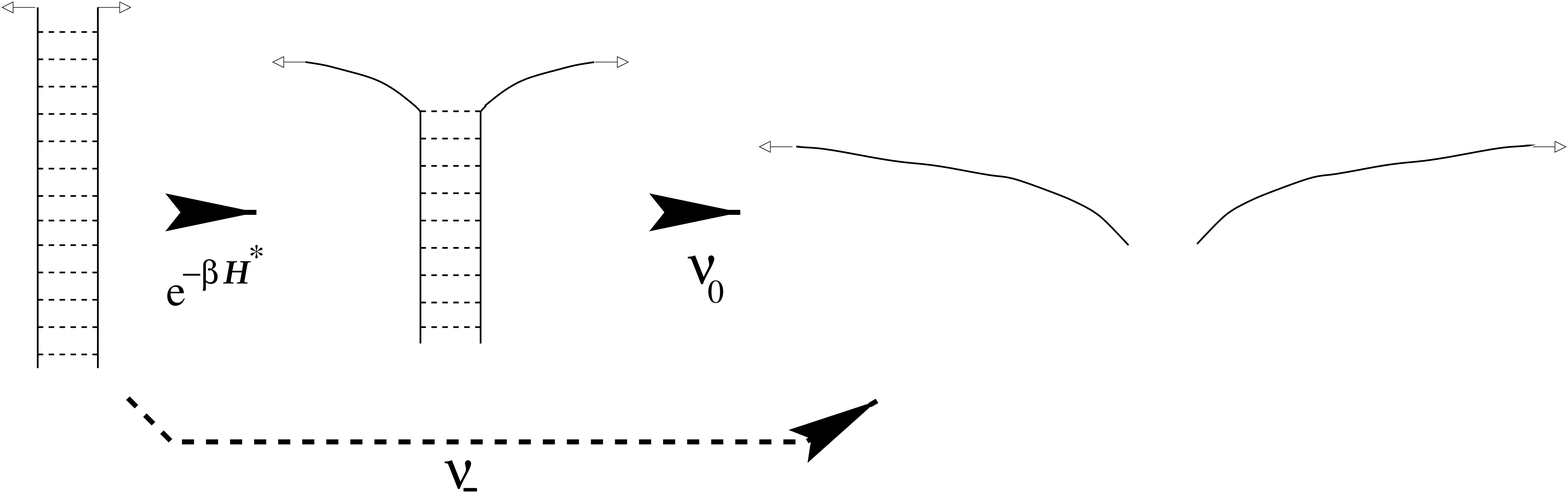}
\vskip 2cm
%\hskip 20cm
\includegraphics[height=130pt,angle=0] {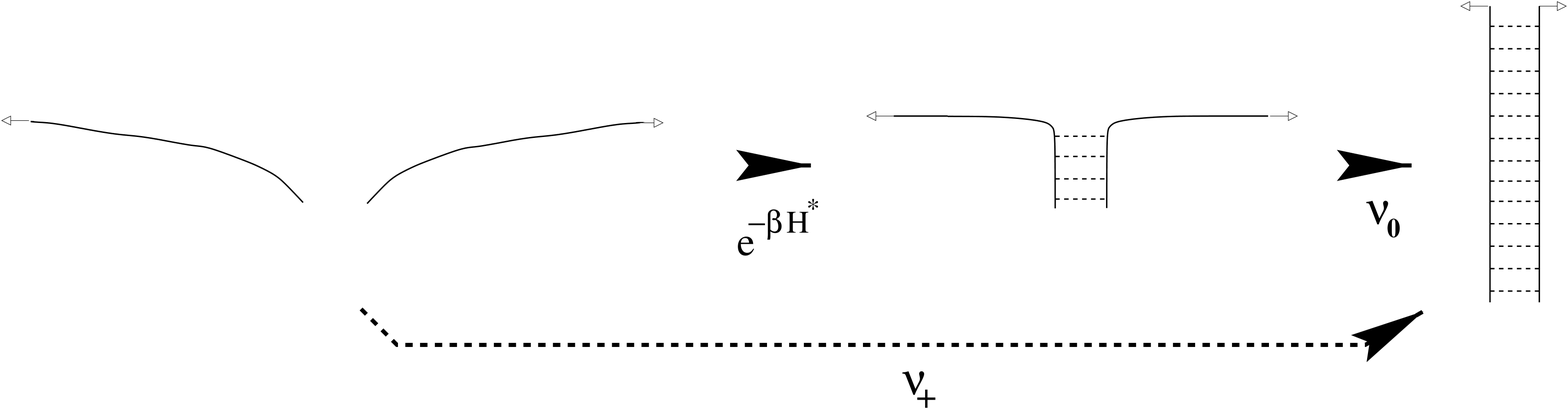}
\begin{flushleft} {\bf B} \end{flushleft}
\vskip 0.5 cm
Cocco. Fig~4. Desired size: 2 columns, height 6.5 cm.
\vskip 2cm
\caption{Transitions states involved in the theoretical calculation of
the kinetic rates. {\bf A} - Unzipping: opening of dsDNA is favorable
at forces $f> f_u$, and the unzipping rate $\nu_-$ is calculated
directly. The nucleation bubble is  of ($\simeq 4$)  base pairs, weakly
 depending  on the force.
{\bf B} - Annealing: when $f<f_u$, dsDNA is
thermodynamically stable; the dissociation rate $\nu _-$ is obtained
indirectly through the calculation of the annealing rate $\nu _+$ of
the metastable ssDNA, $\nu _-= \nu_+ e^{- N \Delta g (f)/k_B T},$
where $\Delta g(f)>0$ is the excess of free-energy per bp of ssDNA
with respect to dsDNA.  The nucleation bubble is of ($\simeq 2$) 
base pairs.}
\label{dessinbulle}
\end{figure}
\end{center}

\begin{center}
\begin{figure}
\includegraphics[height=350pt,angle=-90] {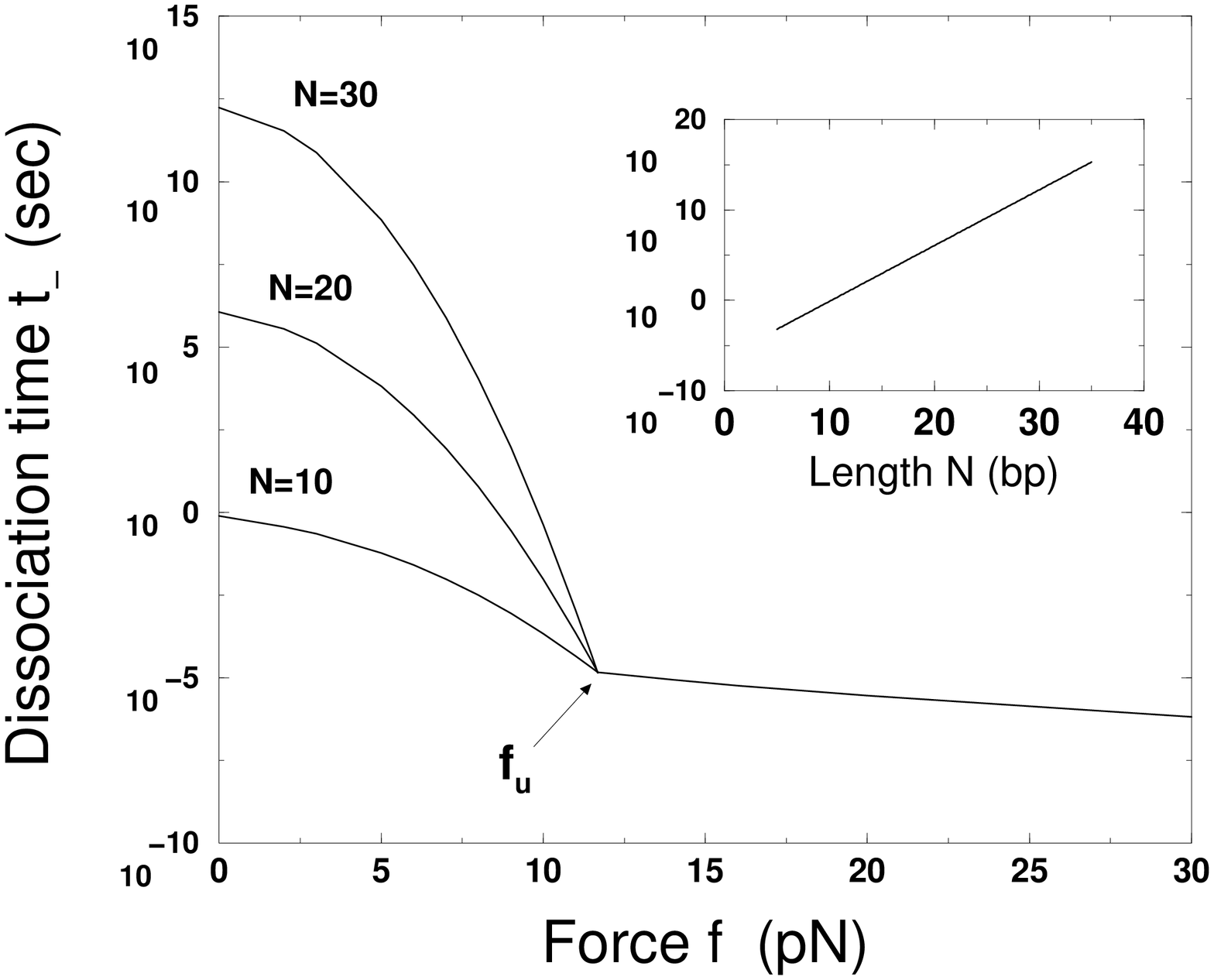}
\vskip 0.5 cm
Cocco. Fig~5. Desired size: 1 column, height 8.5 cm.
\vskip 1.5 cm
\caption{Time of dissociation $t_-$ (in sec) as a
function of the force $f$ (in pN). Three regimes can be distinguished.
For $f<f_u=12$~pN,
the dissociation times depend on the length $N$ of the sequence
($N=10,20,30$~bp from bottom to top). For $f_u<f< f_b=230$~pN, 
the dissociation time is length independent and decreases, as
the energetic barrier to overcome, lowers. For
$f>f_b$, no barrier is left and dissociation is immediate. 
The slope of the logarithm of $t_-$ near $f_u$ is 
$d \log t_{-}/d f = - 8$ \AA~$(f>f_u)$,    
$-2 d_u N +31$ \AA~with $d_u = 5$\AA~$(f<f_u)$. 
Inset: Time of thermal dissociation $t_-$
(for zero force) as a function of the number of base pairs $N$.}
\label{temps}
\label{tempszero}
\end{figure}
\end{center}

\newpage
\begin{center}
\begin{figure}
\includegraphics[height=350pt,angle=-90] {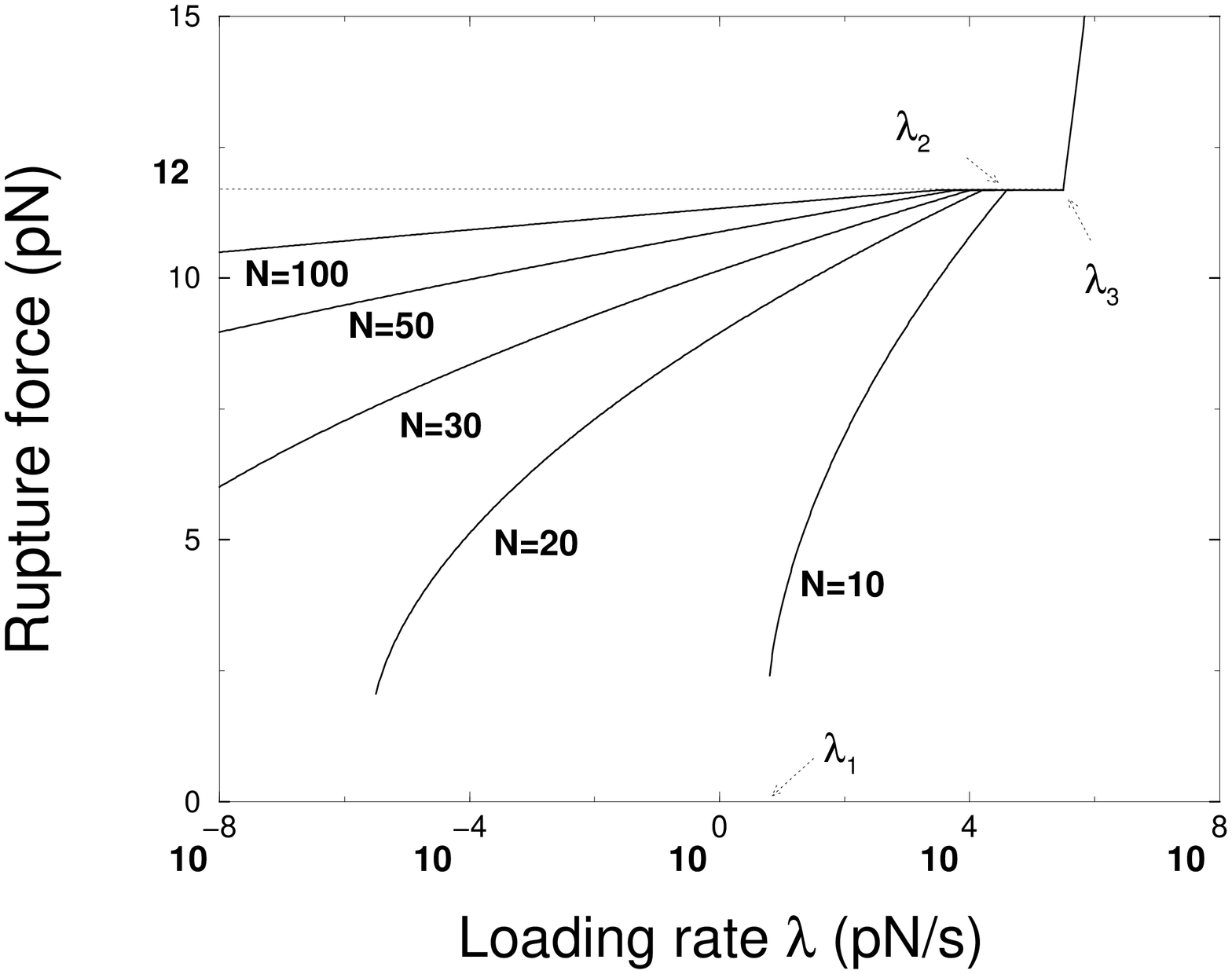}
\vskip 0.5 cm
Cocco. Fig~6. Desired size: 1 column, height 8.5 cm.
\vskip 1.5cm
\caption{Rupture force (pN) as a function of
the loading rate $\lambda$ (pN/sec)  for
five different molecule lengths $N$=10, 20, 30, 50 and 100.
Arrows indicate the different critical loading rates for $N=10$.
Below $\lambda_1$ (=$10^{\,0.8}$ for $N=10$), rupture occurs at essentially
zero force through thermal dissociation. 
For loading rates ranging from  $\lambda _1$ up to $\lambda_2$
(=$10^{\,4.6}$ for $N=10$), the rupture
force is finite, and thermal tunneling is responsible 
for the strong dependence 
on $N$, until the force reaches the equilibrium
value $f_u= 12$~pN. For larger loading rates, the rupture force
is length independent. It increases again as $\lambda > \lambda _3 = 
10^{\,5.5}$~pN/sec,
since the molecule is unable to respond to the force before it becomes very large.}
\label{loading}
\end{figure}
\end{center}


\begin{thebibliography}{99}

\bibitem{Ess}
Essevaz-Roulet, B., Bockelmann, U. \& Heslot, F. (1997) {\em 
Proc. Natl. Acad. Sci. USA} {\bf 94}, 11935-11940;\\
Bockelmann, U., Essevaz-Roulet, B. \& Heslot, F. (1998) {\em
Phys. Rev. E} {\bf 58}, 2386-2394.

\bibitem{Leg}
Leger, J.F., Robert, J., Bourdieu, L., Chatenay, D. \& Marko, J.F.
(1998) {\em Proc. Natl. Acad. Sci. USA}  {\bf 95}, 12295-12299.

\bibitem{Lee}
Lee, G.U., Chrisey, L.A. \& Colton, R.J. (1994) {\em Science} {\bf 266}, 
771-773.

\bibitem{Rie}
Rief, M., Clausen-Schaumann, H. \& Gaub, H.E. (1999)
{\em Nat. Struct. Biol.} {\bf 6}, 346-349.

\bibitem{Stru}
Strunz, T., Oroszlan, K., Schafer, R. \& Guntherodt, H.J. (1999) {\em 
Proc. Natl. Acad. Sci. USA}  {\bf 96}, 11277-11282.

\bibitem{Str}
Strick, T.R., Bensimon, D. \& Croquette, V. (1999) {\em Genetica}
{\bf 106}, 57-62.

\bibitem{Bon98}
Bonnet, G,  Krichevsky, O. \& Libchaber, A. (1998)
{\em Proc. Natl. Acad. Sci. USA} {\bf 95}, 8602-8606.

\bibitem{Thom}
Thompson, R.E. \& Siggia, E.D. (1995) 
{\em Europhys. Lett.} {\bf 31}, 335-340.

\bibitem{Lub}
Lubensky, D.K. \& Nelson, D.R. (2000)
{\em Phys. Rev. Lett.} {\bf 85}, 1572-1575.

\bibitem{Cac}
Bhattacharjee, S.M. (2000) {\em J. Phys. A.} {\bf 33}, L423-L428.

\bibitem{Pro}
Prohofsky, E. (1995)
{\em Statistical mechanics and stability of macromolecules.}
(Cambridge University Press, Cambridge).   

\bibitem{Pey}
Dauxois, T. \& Peyrard, M. (1995) {\em Phys. Rev. E} 
{\bf 51}, 4027-4040. 

\bibitem{Cul}
Cule, D. \& Hwa, T. (1997) {\em Phys. Rev. Lett.} {\bf 79}, 2375-2378.

\bibitem{Coc99}
Cocco, S. \& Monasson, R. (1999) {\em Phys. Rev. Lett.} {\bf 83}, 5178-5181. 
Note that in this previous version the base pair index is discrete.
The passage to the continuous and the definition of $R_1$ will be 
exposed in detail in a forthcoming work.

\bibitem{Coc00}
Cocco, S. \& Monasson, R. (2000) {\em J. Chem. Phys.} {\bf 112}, 10017-10033.  

\bibitem{Lan69}
Langer, J.S. (1969) {\em Ann. Phys. (N.Y.)} {\bf 34}, 258-275. 

\bibitem{Por}
P\"orschke, D. (1971) {\em J. Mol. Biol.} {\bf 62}, 361-381.       

\bibitem{Eva97}
Evans, E. \& Ritchie, K. (1997) {\em Biophys. J.} {\bf 72}, 1541-1555.

\bibitem{Seb00}
Sebastian, K.L. (2000) {\em Phys. Rev. E} {\bf 62}, 1128-1132.

\bibitem{bus}
Smith, S.B., Cui, Y. \& Bustamante, C. (1996) 
{\em Science} {\bf 271}, 795--799;
Bustamante, C., Smith, S.B., Liphardt, J. \& Smith, D.
(2000) {\em Current Opinion in Struct. Biol.} {\bf 10}, 279--285.


\bibitem{deg}
De Gennes, P.G. (1985),  Scaling concepts in polymer physics, 
Cornell University Press.

% strongly reflects the presence 
%of secondary structures 

\bibitem{Maier}
Maier, B., Bensimon, D. \& Croquette, V. (2000)
{\em Proc. Natl. Acad. Sci. USA} {\bf 97}, 12002--12007.

%When dsDNA is unzipped at forces $\simeq 10$~pN, secondary structures of 
%ssDNA are  essentially absent, and the Kuhn length $d$ of the
%bare ssDNA is expected to be shorter, and closer to the backbone 
%elementary length. 

\bibitem{Breslauer}
Breslauer K.J., Frank, R., Blocker, H., \& Marky, L.A.
(1986) {\em Proc. Natl. Acad. Sci. USA} {\bf 83}, 3746--3750.

\bibitem{Ura81}
Urabe, H. \& Tominaga, Y. (1981) {\em J. Phys. Soc. Japan} {\bf 50},
3543-3544. 

\bibitem{Mor}
Morse, P.M. (1929) {\em Phys. Rev.} {\bf 34}, 57-64.

\bibitem{Boh}
Bohm, D. (1951) {\em Quantum Theory}, Prentice-Hall physics series
(Prentice-Hall, New-York). 

 
\end{thebibliography}
\end{document}